\documentclass[3p,times,procedia]{elsarticle}
\usepackage{nupha_ecrc}
 \usepackage{amsmath}
\usepackage{graphicx}
\usepackage{color}

\volume{00}

\firstpage{1}

\journalname{Nuclear Physics A}

\runauth{M.~Nahrgang}

\jid{nupha}

\jnltitlelogo{Nuclear Physics A}

\usepackage{amssymb}

\usepackage[figuresright]{rotating}

\begin{document}

\begin{frontmatter}



\dochead{}

\title{The QCD Critical Point and Related Observables}


\author{Marlene Nahrgang}

\address{Department of Physics, Duke University, Durham, North Carolina 27708-0305, USA}

\begin{abstract}
The search for the critical point of QCD in heavy-ion collision experiments has sparked enormous interest with the completion of phase I of the RHIC beam energy scan. Here, I review the basics of the thermodynamics of the QCD phase transition and its implications for experimental multiplicity fluctuations in heavy-ion collisions. Several sources of noncritical fluctuations impact the observables and need to be understood in addition to the critical phenomena. Recent progress has been made in dynamical modeling of critical fluctuations, which ultimately is indispensable to understand potential signals of the QCD critical point in heavy-ion collision.
\end{abstract}


\end{frontmatter}


\section{The QCD phase diagram and heavy-ion collisions}
\label{sec:intro}
One of the most fascinating opportunities of heavy-ion collisions is the investigation of the phases of strongly interacting matter at finite temperatures and densities. At ultrarelativistic beam energies the interaction of the initial nuclei excites the vacuum and high energies are deposited in the reaction zone, creating a very hot medium of vanishing net-baryon density. At lower beam energies the incoming baryon currents are stopped and the medium is less hot but more dense than at highest beam energies. In these extreme regimes it is expected that a new state of matter, the quark-gluon plasma of colored partonic degrees of freedom is created. The success of models like statistical hadronization \cite{Andronic:2011yq} and of fluid dynamical simulations \cite{Song:2010mg,Schenke:2010rr,Gale:2012rq} in comparison to a variety of experimental data, indicates that the medium thermalizes locally and a connection to QCD thermodynamics is possible. 

Experimental efforts started at lower beam energies, from Bevalac and AGS to the CERN-SPS, mounting up to highest beam energies at RHIC and the LHC. Recent interest has focused on the possibility of scanning the QCD phase diagram by varying the beam energy in the lower energy region. The recent beam energy scan phase I has been completed at RHIC where ample data has been taken and analyzed \cite{Kumar:2012fb,Adamczyk:2013dal,Adamczyk:2014fia,Adare:2015aqk}. Currently, the beam energy and system size scan with the NA61 experiment at the CERN-SPS has started and future programs include the beam energy scan phase II at RHIC and upcoming facilities like the CBM \cite{CBMphysicsbook} experiment at FAIR, GSI Helmholtz Center and NICA \cite{NICAwhitepaper} at JINR in Dubna. The main goal of these endeavors is the investigation of the QCD phase transition and the discovery of a conjectured critical point at finite net-baryon density.

From lattice QCD calculations at zero baryo-chemical potential it is known that the phase change between a partonic and a hadronic medium is an analytic crossover \cite{Aoki:2006we} in the temperature range of $T_c=145-165$~MeV depending on the respective thermodynamic quantity \cite{Borsanyi:2010bp}. From universality considerations \cite{Hatta:2002sj}, studies of effective models \cite{Scavenius:2000qd,Schaefer:2004en,Ratti:2005jh,Schaefer:2007pw,Pawlowski:2005xe,Skokov:2010wb,Herbst:2010rf} and the nonperturbative Dyson-Schwinger approach \cite{Fischer:2012vc,Fischer:2014ata} there are good arguments for the existence of a critical point and a subsequent line of first-order phase transition in the phase diagram at finite net-baryon density. The location of the critical point is strongly parameter-dependent, e.g. if a vector-coupling is included, and on the approach to solve the thermodynamics of the model. 

In this overview, I will focus on the connection between the thermodynamics of QCD matter and the experimental data obtained from heavy-ion collisions. Thermodynamics deals with infinite, static and homogeneous systems, which reach equilibrium in the long-time limit according to a set of controlled thermodynamic variables. Systems created in heavy-ion collisions, however, are very dynamical and undergo a variety of stages, e.g. initial state, pre-equilibrium, quark-gluon plasma and the hadronic gas, each contributing various effects to the final particle spectra. It is therefore necessary to include dynamical effects of the phase transition in realistic modeling of heavy-ion collisions and identify further contributions that impact signals of the QCD critical point.

\section{Fluctuation observables at the phase transition}
\label{sec:observables}
At the phase transition thermodynamic quantities change characteristically. This is seen in lattice calculations \cite{Borsanyi:2013bia,Bazavov:2014noa} by the increase of, for example, the energy density scaled by the temperature, $e/T^4$, due to the liberation of color degrees of freedom. The speed of sound, $c_s^2=(\partial p/\partial e)_S$, has a minimum around the crossover transition and vanishes at a first-order phase transition. Similarly, the compressibility, $\kappa_S=-1/V(\partial V/\partial p)_S$, has a maximum around the crossover transition and diverges at the first-order phase transition. These quantities are related to an anomaly in the pressure. It is straight forward to describe the phase transition in heavy-ion collisions via fluid dynamical calculations on the level of the equation of state and transport coefficients. In early (ideal) fluid dynamical calculations, for example, a pronounced minimum in the slope of the directed flow $v_1$ was observed at a first-order phase transition compared to the absence of a phase transition \cite{Stoecker:2004qu}. Since modern calculations \cite{Steinheimer:2014pfa} with improved schemes for particle production find a substantial reduction of this signal and hardly see any difference between the equation of state of a crossover and of a first-order phase transition, there is currently no confirmed impact of the anomaly in the pressure in simulations of heavy-ion collisions. 

\begin{figure}
\includegraphics[width=0.19\textwidth]{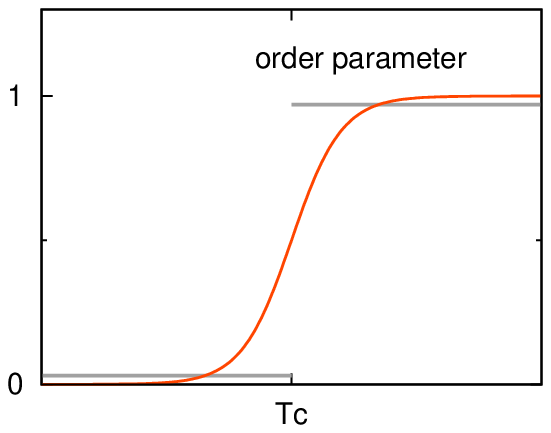}
\includegraphics[width=0.19\textwidth]{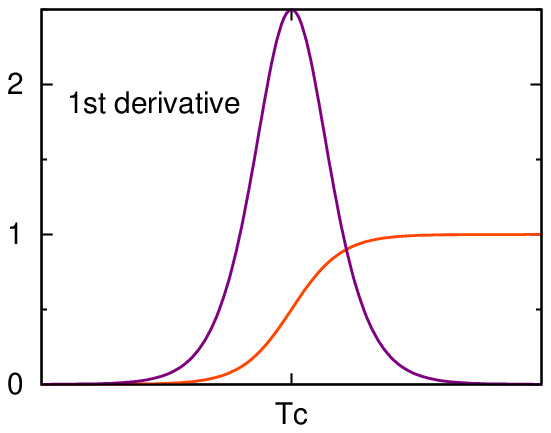}
\includegraphics[width=0.19\textwidth]{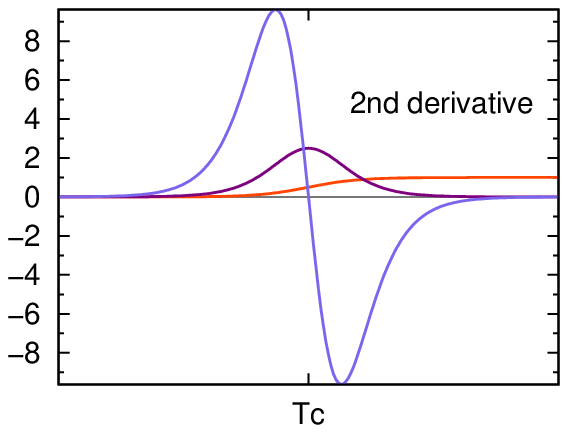}
\includegraphics[width=0.19\textwidth]{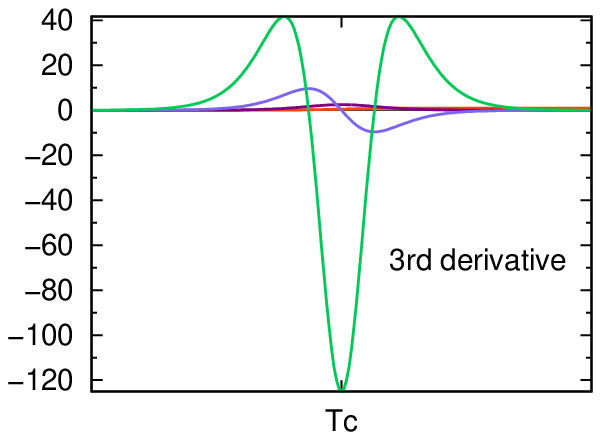}
\includegraphics[width=0.19\textwidth]{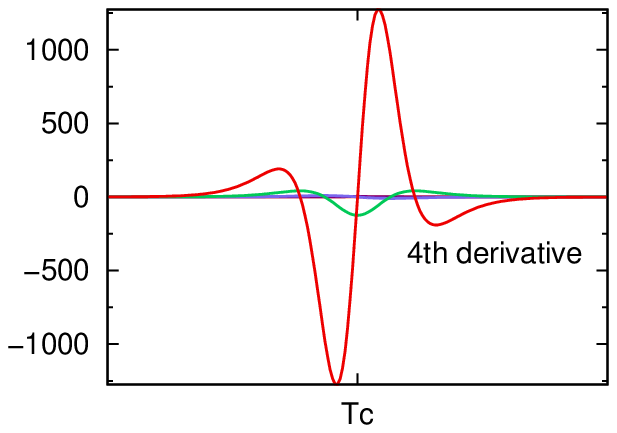}
\caption{Sketch of the order parameter and its derivatives at a phase transition.}
\label{fig:derivatives}
\end{figure}

\subsection{Susceptibilities and event-by-event fluctuations}
While the order parameter(s) change characteristically at the phase transition, more details are revealed by taking derivatives, see Fig.~\ref{fig:derivatives} for illustration. Derivatives of thermodynamic quantities are related to ensemble fluctuations, which can be measured as event-by-event fluctuations in heavy-ion collisions. The quantities to look at in this respect are the susceptibilities, which are defined as derivatives of the pressure with respect to the chemical potential
\begin{equation}
 \chi_n=\frac{\partial^n(P/T^4)}{\partial(\mu/T)^n}\, .
\end{equation}
Starting the derivation from the definition of the grandcanonical partition function one can show that the susceptibilities relate to cumulants of the event-by-event multiplicity distributions via
 \begin{equation}
 \chi_1=\frac{1}{VT^3} \langle N\rangle\, , \quad  \chi_2=\frac{1}{VT^3} \langle (\Delta N)^2\rangle\, ,\quad  \chi_3=\frac{1}{VT^3} \langle (\Delta N)^3\rangle\, , 
\end{equation}
 \begin{equation}
 \chi_4=\frac{1}{VT^3}\langle(\Delta N)^4\rangle_c \equiv\frac{1}{VT^3} \left(\langle (\Delta N)^4\rangle-3\langle (\Delta N)^2\rangle^2\right)\, .
\end{equation}
Here, $N$ is the number of the measured particles or conserved charges and $\Delta N=N-\langle N\rangle$ is the fluctuation around the event-averaged mean. Under the assumption that fluctuations are determined at a hypersurface which corresponds to a fixed temperature, the ratios of susceptibilities 
\begin{equation}
\frac{\chi_2}{\chi_1}=\frac{{\sigma^2}}{M}\, ,\quad\quad  \quad \quad \frac{\chi_3}{\chi_2}={S}\sigma\, ,\quad \quad\quad  \quad \frac{\chi_4}{\chi_2}={\kappa}\sigma^2
\label{eq:ratios}
\end{equation}
are independent of the volume to zero-th order in volume fluctuations.
The mean $M$, the variance $\sigma^2$, the skewness $S$ and the kurtosis $\kappa$ are obtained from measured event-by-event multiplicity distributions and can via Eqs. (\ref{eq:ratios}) be compared to thermodynamic calculations.

\subsection{Fluctuations at a critical point}
The expected signal in fluctuation observables depends on the type of phase transition, crossover, critical point or first-order phase transition. In particular, the features of a critical point stand out. Due to the divergence of the correlation length of fluctuations, $\xi\to\infty$, microscopic details of the specific interaction become less and less important the closer a system approaches its critical point. The relevant physics are determined by only a couple of parameters, such as the dimensionality of the system and other scale-invariant quantities. In this sense, large classes of physical phenomena, exhibit the same universal behavior in the scaling regime close to a critical point. QCD with finite quark-masses belongs to the universality class of $3$d Ising models. 
In static and equilibrated systems, critical phenomena emerge as a result of the diverging correlations, most prominently the divergence of the fluctuations of the critical mode $\sigma$. It has been shown that the higher-order cumulants are more sensitive to the divergence of the correlation 
  \begin{equation}
   \langle\Delta\sigma^2\rangle\propto\xi^2\, , \qquad \langle\Delta\sigma^3\rangle\propto\xi^{9/2}\, , \qquad 
   \langle\Delta\sigma^4\rangle_c\propto\xi^7\, ,
   \label{eq:sigmaflucs}
  \end{equation}
where the $\xi$ dependence of the couplings from the $3$d Ising universality class has been employed \cite{Stephanov:2008qz}.

The growth of the correlation length takes finite time, and close to the critical point not only the fluctuations diverge as a function of the correlation length with some critical exponent, but so do associated relaxation times $\tau_{\rm rel}\propto \xi^{\, z}$, where $z$ is a dynamical critical exponent \cite{Hohenberg:1977ym}. This implies that in a dynamical setup, a system which traverses the critical point in any finite amount of time is necessarily driven out of equilibrium, even if it reached thermal equilibrium in some state away from the critical point. This effect is called critical slowing down and known to weaken the phenomena of criticality. A phenomenological evolution equation of the correlation length has been studied in \cite{Berdnikov:1999ph} and the maximal size of the correlation length was found to extend to $1.5-2.5$~fm.

Low-energy effective models of QCD, like the quark-meson (QM) or Polyakov quark-meson model (PQM) include some aspects of the QCD phase transition, like the chiral $O(4)$ or the $Z(2)$ symmetry. Compared to lattice calculations they do not capture all the aspects of QCD but can easily be studied at finite $\mu_B$, give an insight into the active degrees of freedom and real-time calculations can in principle be carried out.  
Typically, mean-field critical exponents and critical phenomena do not describe the correct physics as mesonic fluctuations and thus important correlations are neglected. Functional renormalization group methods can go beyond the mean-field approximation and provide valuable information about the fluctuations in conserved-charge densities. In \cite{Skokov:2012ds} it has been shown that $\chi_4/\chi_2$ has a prominent $T$-$\mu_B$-dependence and the signal grows stronger as the critical point is approached.

\subsection{Fluctuations at a first-order phase transition}
Fluctuations at a first-order phase transition are of a very different nature. They occur only if a nonequilibrium situation is present. At the transition temperature the high and low temperature phase coexist. These two stable phases are separated by a potential barrier, the latent heat. At temperatures above and below $T_c$ metastable states exist as remnants of the phase coexistence, still separated by a barrier. Thus, a system which is in thermal equilibrium above the phase transition and cools to a temperature slightly below the phase transition needs a finite time to relax over this barrier into the new stable phase. If this nucleation rate is small and the expansion (cooling) is fast, parts of the system remain in the metastable phase until it becomes unstable at the lower spinodal. Here, the low-momentum modes are amplified by spinodal decomposition, which leads to domain formation of the two phases \cite{Csernai:1995zn,Mishustin:1998eq,Scavenius:2000bb,Keranen:2002sw,Randrup:2009gp,Randrup:2010ax,Steinheimer:2012gc,Herold:2013qda}.
 
 \subsection{Critical fluctuations in heavy-ion collisions}
 In order to obtain predictions for measurable particles, the work in \cite{Stephanov:1998dy,Stephanov:1999zu,Stephanov:2008qz,Athanasiou:2010kw} applied the theory of critical phenomena to heavy-ion collisions by looking at a coupling of the order parameter $\sigma$ to protons via $g_p \bar{p}\sigma p$ and derived expressions for the $n$-th order cumulants of proton multiplicity distributions including the leading-order critical contribution
 \begin{equation}
  \langle\Delta N_p^n\rangle =  \langle (\Delta N_p^0)^n\rangle_c  + \,\langle (V\Delta\sigma)^n\rangle_c \left(\frac{-g_{p}d_p}{T}\int_k \frac{n_p^0(k)(1-n_p^0(k))}{\gamma_p(k)}\right)^n\, ,
  \label{eq:hrgflucs}
\end{equation}
where $\langle (\Delta N_p^0)^n\rangle_c$ is the $n$-order cumulant of the noncritical baseline, e.g. for a Poisson distribution, $n_p^0(k)$ is the Fermi-Dirac thermal distribution of protons and $\gamma_p(k)=\sqrt{k^2+m_p^2}$. Estimations based on these formulas have been the main motivation to search for the QCD critical point in heavy-ion collision experiments.
 
\begin{figure}
  \includegraphics[width=0.5\textwidth]{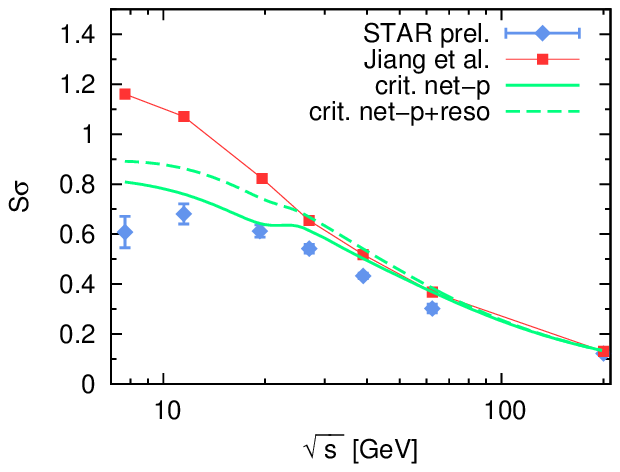}
  \includegraphics[width=0.5\textwidth]{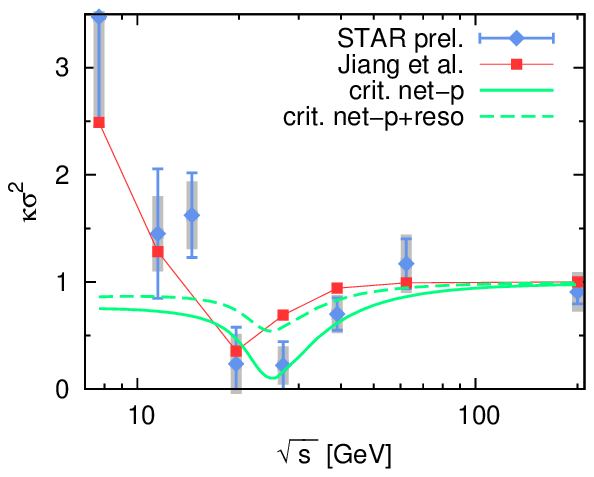}
  \caption{Net-proton fluctuations as a function of beam energy compared to preliminary data from STAR \cite{Luo:2015ewa} ($|y|<0.5$, $0.4<p_T<2$~(GeV/c)). The light green curves are based on Eq. (\ref{eq:hrgflucs}) with fixed parameters and the dashed line includes fluctuations from resonance decays \cite{workinprogress}. The red squares are obtained from particlization over a fluid dynamical hypersurface with parameters tuned to reproduce the experimental data for the cumulants \cite{Jiang:2015hri}. In all curves criticality is introduced via the coupling of (anti-)protons to static, equilibrium $\sigma$ fluctuations \cite{Stephanov:1998dy,Stephanov:1999zu,Stephanov:2008qz,Athanasiou:2010kw}.}
  \label{fig:netpflucs}
 \end{figure}

 Net-proton fluctuations based on Eq. (\ref{eq:hrgflucs}) are shown in Fig.  \ref{fig:netpflucs} (green, solid lines) as a function of the beam energy in comparison with preliminary data from the STAR collaboration \cite{Luo:2015ewa}. The results depend strongly on the correlation length, for which an assumption according to \cite{Athanasiou:2010kw} is used with $\xi_{\rm max}=2.6$~fm at $\mu_B=0.15$~GeV and width $0.4$~GeV. Furthermore, the  $3$d Ising model parameters are fixed to $\tilde\lambda_3=1.5$ and $\tilde\lambda_4=13$, and the coupling is $g_p=9$. Freeze-out conditions are taken from \cite{Cleymans:2005xv}. The experimental cuts have been included in the thermal integrals.
 Recently, this approach has been extended to correlated particle production over a hypersurface of a fluid dynamical expansion \cite{Jiang:2015hri,Jiang:2015cnt}. Here, the mentioned parameters have been tuned for each collision energy to rougly reproduce the experimental data for the cumulants separately. The results shown here (red squares)\footnote{L.~Jiang, P.~Li and H.~Song, private communication} are for parameter set I in \cite{Jiang:2015hri}. The increasing trend at low energies in $\kappa\sigma^2$ can be described with appropriate parameter choices from a particlization over a fluid dynamic hypersurface, but the decreasing trend at the lower energies in $S\sigma$ seems to be difficult to obtain. 
 None of these models include any dynamical evolution of the critical fluctuations.
 
 \section{Non-critical sources of fluctuations}
 \label{sec:noncritical}
 Fluctuations reveal more details of the phase transition, but also of everything else that affects the multiplicity distributions. In order to separate the interesting fluctuations from the the non-critical contributions it is therefore indispensable to carefully identify and study further sources of fluctuations: 
 \begin{itemize}
  \item[\textbullet] If the limited detector acceptance is smaller than the critical correlation length of the measured particles, the signal is diminished. Similarly, any detector efficiency which is smaller than $100$\% due induces additional fluctuations to the finite probability that a certain particle is not detected. This washes out the critical fluctuation signal. Under the assumption that the detection of each particle is independent with constant efficiency, i.e. that the distribution of $n$ detected particles out of $N$ particles reaching the detector in acceptance is binomial, this effect can be corrected according to \cite{Bzdak:2012ab,Bzdak:2013pha}. If the assumption of a binomial distribution is justified remains to be carefully checked by the experiments.
  \item[\textbullet] Experimentally, it is not feasible to detect the neutrons. Quasielastic scatterings of protons or neutrons with pions via a delta resonance ($p(n)+\pi\to\Delta\to n(p)+\pi$) in the hadronic phase wash out the critical fluctuations in net-baryon number. It is similar to limited acceptance or efficiency effects and under the assumption of complete isospin randomization, i.e. that the number of protons among the baryons is given by a binomial distribution, can be corrected \cite{Kitazawa:2011wh,Kitazawa:2012at,Nahrgang:2014fza}.
  \item[\textbullet] A very similar non-critical source of fluctuations is the decay of resonances in the late hadronic stage \cite{Nahrgang:2014fza}. In Fig.~2 the impact of the stochastic contribution from resonance decay to the net-proton fluctuations is shown by the green, dashed line. The parameters are the same as described above. It is seen that the critical fluctuations are significantly reduced but survive, in particular in $\chi_4/\chi_2$, when resonance decays are included \cite{workinprogress}.
  \item[\textbullet] Volume fluctuations are inevitable in heavy-ion collision experiments and they are expected to affect all types of collisions, except very central ones \cite{Skokov:2012ds,Sangaline:2015bma}. Recently, a variety of fluctuation observables, which are not only independent of the volume average but also independent of any volume fluctuations, so called strongly intensive measures have been proposed \cite{Gorenstein:2011vq,Sangaline:2015bma}. 
  \item[\textbullet] Toward lower beam energies more of the initial net-baryon number is transported to midrapidity, which means that the assumption of a  grandcanonical ensemble might not be justified any more and the effects of global net-baryon number conservation start to affect the fluctuation observables. In microscopic transport models the individual scatterings are of microcanonical nature and net-baryon number conservation was found to have a huge impact on the net-baryon number kurtosis with the net-proton kurtosis to slightly follow this trend \cite{Schuster:2009jv}. These calculations serve as valuable baseline studies. Starting from a binomial distribution the effects of global net-baryon number conservation have been studied in \cite{Bzdak:2012an} as well.
  \item[\textbullet] Initial fluctuations due to baryon stopping become increasingly important at low beam energies, since most of the measured protons come from the initial baryon current, which is transported to midrapidity during the initial collisions. Several models are able to somewhat different degrees to describe the rapidity distribution of particle production at lower beam energies but the impact of initial baryon-number fluctuations at midrapidity has not yet been studied.
 \end{itemize}
 A recent MC sampling of critical and some noncritical contributions to the variance of pion fluctuations from a similar coupling as discussed for protons above found a large effect of noncritical fluctuations \cite{Hippert:2015rwa}.

  \section{Dynamics of critical fluctuations}
 \label{sec:dynamics}
 The estimations obtained in \cite{Stephanov:1998dy,Stephanov:1999zu,Stephanov:2008qz,Athanasiou:2010kw} and its recent refinements in \cite{Jiang:2015hri} and \cite{workinprogress} as shown in Fig. \ref{fig:netpflucs} are based on static, equilibrium expectations for $\langle(\Delta\sigma)^n\rangle_c$ albeit with a phenomenological input for a finite correlation length. In order to go beyond and achieve a quantitative statement about the connection between the thermodynamics of the QCD critical point and experimental observables the dynamics of fluctuations needs to be included in realistic modeling of heavy-ion collisions. 
  
 In \cite{Mukherjee:2015swa} it is shown that the magnitude and sign of the critical contributions to the fluctuation signals can be different in non-equilibrium situations compared to equilibrium expectations due to the importance of memory effects. In this work, the real-time evolution of non-Gaussian cumulants is studied in the scaling regime, where 
$L_{\rm micro}\ll\xi\ll L_{\rm sys}$ via an expansion of the Fokker-Planck dynamics of the order parameter. In general different combinations of trajectories, chemical freeze-out conditions and relaxation times $\tau_{\rm rel}$ can give similar results for the $\sqrt{s}$-dependence of the higher-order cumulants.
 
 In order to investigate the dynamics of critical fluctuations within a realistic space-time evolution of heavy-ion collisions, the model of nonequilibrium chiral fluid dynamics (N$\chi$FD) has been developed continuously over the recent years \cite{Nahrgang:2011mg,Nahrgang:2011mv,Nahrgang:2011vn,Herold:2013bi,Herold:2013qda,Herold:2014zoa,Herold:2016uvv}. It explicitly propagates the order parameter fields coupled to a fluid dynamical evolution of the QGP. The order parameter of chiral symmetry, the $\sigma$ field evolves via a stochastic relaxation equation
 \begin{equation}
  \partial_\mu\partial^\mu\sigma+\frac{\delta U}{\delta\sigma}+g\rho_s+\eta\partial_t\sigma=\xi\, ,
 \end{equation}
 where the chiral potential $U$, the scalar density $\rho_s$ and the damping coefficient $\eta$ are obtained from a chiral effective model like the (P)QM model. Following the fluctuation-dissipation theorem the magnitude of the noise field $\xi$ is determined by the damping coefficient and taken in the Gaussian approximation. Due to the presence of $\xi$ the evolution of the $\sigma$ field becomes stochastic. The fermionic part of the Lagrangian is treated fluid dynamically and serves as heat bath for the nonequilibrium dynamics of the order parameter. The fluid dynamical evolution and the $\sigma$ field evolution can be consistently derived within the 2PI effective action approach. The two sectors are coupled via a stochastic source term in the fluid dynamical equations
  \begin{equation}
    \partial_\mu T^{\mu\nu}_{\rm q}={ S^\nu=-\partial_\mu T^{\mu\nu}_\sigma}\, ,\quad  \partial_\mu N^{\mu}_{\rm q}= 0\, .
 \end{equation}
 Consequently, the fluid dynamical fields itself become stochastic quantities.
 This approach produces effects of supercooling, reheating and domain formation at the first-order phase transition, as well as critical slowing down at the critical point. Recently, a particlization has been added and the dynamics along a trajectory on the crossover side of the critical point shows that critical contributions to the net-proton kurtosis are present in dynamical systems \cite{Herold:2016uvv}. Further and more quantitative work is underway.
 
 \section{Fluid dynamical fluctuations}
 \label{sec:FDF}
 In N$\chi$FD the fluctuations in the fluid dynamical fields were induced by the stochastic evolution in the $\sigma$ field. At finite baryochemical potential, however, the $\sigma$ field mixes with the net-baryon density, which in the long-time limit becomes the true critical mode due to the diffusive dynamics \cite{Son:2004iv}. The inclusion of fluctuations into the  fluid dynamical evolution equation is therefore a promising route to the dynamical treatment of critical fluctuations. While conventional fluid dynamics propagates only thermal averages of the fluid dynamical fields, we know that already in equilibrium there are thermal fluctuations and that the fast processes, which lead to local equilibration also lead to noise. For a consistent treatment of viscosities, of the baryon conductivity and the dissipation-fluctuation theorem the equations for fluctuating viscous fluid dynamics read \cite{Kapusta:2011gt,Murase:2013tma}
\begin{align}
 T^{\mu\nu}&=T^{\mu\nu}_{\rm eq}+\Delta T^{\mu\nu}_{\rm visc}+\Xi^{\mu\nu} \, ,\\
 N^{\mu}&=N^{\mu}_{\rm eq}+\Delta N^{\mu}_{\rm visc}+{\rm I}^{\mu}\, .
\end{align}
with the autocorrelation of the noise term
\begin{align}
\langle \Xi^{\mu\nu}(x)\Xi^{\alpha\beta}(x')\rangle & =  2T[\eta(\Delta^{\mu\alpha}\Delta^{\nu\beta}+\Delta^{\mu\beta}\Delta^{\nu\alpha})+(\zeta-2/3\eta)\Delta^{\mu\nu}\Delta^{\alpha\beta}]\delta^4(x-x')\, ,\\
\langle {\rm I}^{\mu}(x){\rm I}^{\nu}(x')\rangle & = 2T\kappa\Delta^{\mu\nu}\delta^4(x-x')\, .
\end{align}
Assuming enhanced conductivities at the critical point, the nonlinear fluid dynamical equations produce enhanced correlator of fluid dynamical densities in a  simple Bjorken expansion \cite{Kapusta:2012zb}. First numerical implementations \cite{Young:2014pka}, suggest an enhancement of flow due to the additional fluctuations, crucial questions of the renormalization of viscosities and the equation of state due to nonlinear effects \cite{Chafin:2012eq,Kovtun:2011np} and the reproduction of equilibrium expectations remain and are currently under investigation.
 
 \section{Toward the discovery of the critical point}
 \label{sec:discovery}
 Through a realistic dynamical modeling of heavy-ion collisions the QCD critical point and experimental data can be connected and understood. 
 The uncertainties of dynamical models can be reduced by relying on input from first-priniciple calculations whenever possible. Remaining unknown quantities can be tuned by comparing to experimental observables, which are not signals for the critical point. For noncritical bulk observables this procedure is well established in current fluid dynamical calculations for example. In order to search for the QCD critical point, it becomes necessary to include the dynamics of critical fluctuations and to develop appropriate interfaces of these fluctuations with the initial fluctuations and correlations and for particlization.
  Such a framework is also capable to treat noncritical effects quantitatively. 
  This paper outlined a couple of recent approaches to address the dynamics of critical fluctuations and gave an outlook of important current and future developments, in particular, in form of fluid dynamical fluctuations and their integration into realistic descriptions of the space-time evolution of the quark-gluon plasma and the hadronic stage. With these tools the discovery of the QCD critical point and the investigation of its properties at current and future heavy-ion collision experiments will become possible.

\section*{Acknowledgments}
M.N. acknowledges support from a fellowship within the Postdoc-Program of the German Academic Exchange Service (DAAD). This work was supported by the U.S. department of energy under grant DE-FG02-05ER41367.



\bibliography{biblio_nahrgang}

\begin{thebibliography}{10}
\expandafter\ifx\csname url\endcsname\relax
  \def\url#1{\texttt{#1}}\fi
\expandafter\ifx\csname urlprefix\endcsname\relax\def\urlprefix{URL }\fi
\expandafter\ifx\csname href\endcsname\relax
  \def\href#1#2{#2} \def\path#1{#1}\fi

\bibitem{Andronic:2011yq}
A.~Andronic, P.~Braun-Munzinger, K.~Redlich, J.~Stachel, {The thermal model on
  the verge of the ultimate test: particle production in Pb-Pb collisions at
  the LHC}, J. Phys. G38 (2011) 124081.
\newblock \href {http://arxiv.org/abs/1106.6321} {\path{arXiv:1106.6321}},
  \href {http://dx.doi.org/10.1088/0954-3899/38/12/124081}
  {\path{doi:10.1088/0954-3899/38/12/124081}}.

\bibitem{Song:2010mg}
H.~Song, S.~A. Bass, U.~Heinz, T.~Hirano, C.~Shen, {200 A GeV Au+Au collisions
  serve a nearly perfect quark-gluon liquid}, Phys.Rev.Lett. 106 (2011) 192301.
\newblock \href {http://arxiv.org/abs/1011.2783} {\path{arXiv:1011.2783}},
  \href {http://dx.doi.org/10.1103/PhysRevLett.106.192301,
  10.1103/PhysRevLett.109.139904} {\path{doi:10.1103/PhysRevLett.106.192301,
  10.1103/PhysRevLett.109.139904}}.

\bibitem{Schenke:2010rr}
B.~Schenke, S.~Jeon, C.~Gale, {Elliptic and triangular flow in event-by-event
  (3+1)D viscous hydrodynamics}, Phys.Rev.Lett. 106 (2011) 042301.
\newblock \href {http://arxiv.org/abs/1009.3244} {\path{arXiv:1009.3244}},
  \href {http://dx.doi.org/10.1103/PhysRevLett.106.042301}
  {\path{doi:10.1103/PhysRevLett.106.042301}}.

\bibitem{Gale:2012rq}
C.~Gale, S.~Jeon, B.~Schenke, P.~Tribedy, R.~Venugopalan, {Event-by-event
  anisotropic flow in heavy-ion collisions from combined Yang-Mills and viscous
  fluid dynamics}, Phys.Rev.Lett. 110 (2013) 012302.
\newblock \href {http://arxiv.org/abs/1209.6330} {\path{arXiv:1209.6330}},
  \href {http://dx.doi.org/10.1103/PhysRevLett.110.012302}
  {\path{doi:10.1103/PhysRevLett.110.012302}}.

\bibitem{Kumar:2012fb}
L.~Kumar, {STAR Results from the RHIC Beam Energy Scan-I}, Nucl. Phys. A904-905
  (2013) 256c--263c.
\newblock \href {http://arxiv.org/abs/1211.1350} {\path{arXiv:1211.1350}},
  \href {http://dx.doi.org/10.1016/j.nuclphysa.2013.01.070}
  {\path{doi:10.1016/j.nuclphysa.2013.01.070}}.

\bibitem{Adamczyk:2013dal}
L.~Adamczyk, et~al., {Energy Dependence of Moments of Net-proton Multiplicity
  Distributions at RHIC}, Phys.Rev.Lett. 112~(3) (2014) 032302.
\newblock \href {http://arxiv.org/abs/1309.5681} {\path{arXiv:1309.5681}},
  \href {http://dx.doi.org/10.1103/PhysRevLett.112.032302}
  {\path{doi:10.1103/PhysRevLett.112.032302}}.

\bibitem{Adamczyk:2014fia}
L.~Adamczyk, et~al., {Beam energy dependence of moments of the net-charge
  multiplicity distributions in Au+Au collisions at RHIC}, Phys.Rev.Lett. 113
  (2014) 092301.
\newblock \href {http://arxiv.org/abs/1402.1558} {\path{arXiv:1402.1558}},
  \href {http://dx.doi.org/10.1103/PhysRevLett.113.092301}
  {\path{doi:10.1103/PhysRevLett.113.092301}}.

\bibitem{Adare:2015aqk}
A.~Adare, et~al., {Measurement of higher cumulants of net-charge multiplicity
  distributions in Au$+$Au collisions at $\sqrt{s_{_{NN}}}=7.7-200$ GeV}\href
  {http://arxiv.org/abs/1506.07834} {\path{arXiv:1506.07834}}.

\bibitem{CBMphysicsbook}
B.~Friman, C.~Hohne, J.~Knoll, S.~Leupold, J.~Randrup, R.~Rapp, P.~Senger, {The
  CBM physics book: Compressed baryonic matter in laboratory experiments},
  Lect. Notes Phys. 814 (2011) pp. 980.
\newblock \href {http://dx.doi.org/10.1007/978-3-642-13293-3}
  {\path{doi:10.1007/978-3-642-13293-3}}.

\bibitem{NICAwhitepaper}
\href{http://theor0.jinr.ru/twiki-cgi/view/NICA/NICAWhitePaper}{[link]}.
\newline\urlprefix\url{http://theor0.jinr.ru/twiki-cgi/view/NICA/NICAWhitePaper}

\bibitem{Aoki:2006we}
Y.~Aoki, G.~Endrodi, Z.~Fodor, S.~D. Katz, K.~K. Szabo, {The Order of the
  quantum chromodynamics transition predicted by the standard model of particle
  physics}, Nature 443 (2006) 675--678.
\newblock \href {http://arxiv.org/abs/hep-lat/0611014}
  {\path{arXiv:hep-lat/0611014}}, \href {http://dx.doi.org/10.1038/nature05120}
  {\path{doi:10.1038/nature05120}}.

\bibitem{Borsanyi:2010bp}
S.~Borsanyi, Z.~Fodor, C.~Hoelbling, S.~D. Katz, S.~Krieg, C.~Ratti, K.~K.
  Szabo, {Is there still any $T_c$ mystery in lattice QCD? Results with
  physical masses in the continuum limit III}, JHEP 09 (2010) 073.
\newblock \href {http://arxiv.org/abs/1005.3508} {\path{arXiv:1005.3508}},
  \href {http://dx.doi.org/10.1007/JHEP09(2010)073}
  {\path{doi:10.1007/JHEP09(2010)073}}.

\bibitem{Hatta:2002sj}
Y.~Hatta, T.~Ikeda, {Universality, the QCD critical / tricritical point and the
  quark number susceptibility}, Phys.Rev. D67 (2003) 014028.
\newblock \href {http://arxiv.org/abs/hep-ph/0210284}
  {\path{arXiv:hep-ph/0210284}}, \href
  {http://dx.doi.org/10.1103/PhysRevD.67.014028}
  {\path{doi:10.1103/PhysRevD.67.014028}}.

\bibitem{Scavenius:2000qd}
O.~Scavenius, A.~Mocsy, I.~N. Mishustin, D.~H. Rischke, {Chiral phase
  transition within effective models with constituent quarks}, Phys. Rev. C64
  (2001) 045202.
\newblock \href {http://arxiv.org/abs/nucl-th/0007030}
  {\path{arXiv:nucl-th/0007030}}, \href
  {http://dx.doi.org/10.1103/PhysRevC.64.045202}
  {\path{doi:10.1103/PhysRevC.64.045202}}.

\bibitem{Schaefer:2004en}
B.-J. Schaefer, J.~Wambach, {The Phase diagram of the quark meson model}, Nucl.
  Phys. A757 (2005) 479--492.
\newblock \href {http://arxiv.org/abs/nucl-th/0403039}
  {\path{arXiv:nucl-th/0403039}}, \href
  {http://dx.doi.org/10.1016/j.nuclphysa.2005.04.012}
  {\path{doi:10.1016/j.nuclphysa.2005.04.012}}.

\bibitem{Ratti:2005jh}
C.~Ratti, M.~A. Thaler, W.~Weise, {Phases of QCD: Lattice thermodynamics and a
  field theoretical model}, Phys. Rev. D73 (2006) 014019.
\newblock \href {http://arxiv.org/abs/hep-ph/0506234}
  {\path{arXiv:hep-ph/0506234}}, \href
  {http://dx.doi.org/10.1103/PhysRevD.73.014019}
  {\path{doi:10.1103/PhysRevD.73.014019}}.

\bibitem{Schaefer:2007pw}
B.-J. Schaefer, J.~M. Pawlowski, J.~Wambach, {The Phase Structure of the
  Polyakov--Quark-Meson Model}, Phys.Rev. D76 (2007) 074023.
\newblock \href {http://arxiv.org/abs/0704.3234} {\path{arXiv:0704.3234}},
  \href {http://dx.doi.org/10.1103/PhysRevD.76.074023}
  {\path{doi:10.1103/PhysRevD.76.074023}}.

\bibitem{Pawlowski:2005xe}
J.~M. Pawlowski, {Aspects of the functional renormalisation group}, Annals
  Phys. 322 (2007) 2831--2915.
\newblock \href {http://arxiv.org/abs/hep-th/0512261}
  {\path{arXiv:hep-th/0512261}}, \href
  {http://dx.doi.org/10.1016/j.aop.2007.01.007}
  {\path{doi:10.1016/j.aop.2007.01.007}}.

\bibitem{Skokov:2010wb}
V.~Skokov, B.~Stokic, B.~Friman, K.~Redlich, {Meson fluctuations and
  thermodynamics of the Polyakov loop extended quark-meson model}, Phys. Rev.
  C82 (2010) 015206.
\newblock \href {http://arxiv.org/abs/1004.2665} {\path{arXiv:1004.2665}},
  \href {http://dx.doi.org/10.1103/PhysRevC.82.015206}
  {\path{doi:10.1103/PhysRevC.82.015206}}.

\bibitem{Herbst:2010rf}
T.~K. Herbst, J.~M. Pawlowski, B.-J. Schaefer, {The phase structure of the
  Polyakov--quark-meson model beyond mean field}, Phys.Lett. B696 (2011)
  58--67.
\newblock \href {http://arxiv.org/abs/1008.0081} {\path{arXiv:1008.0081}},
  \href {http://dx.doi.org/10.1016/j.physletb.2010.12.003}
  {\path{doi:10.1016/j.physletb.2010.12.003}}.

\bibitem{Fischer:2012vc}
C.~S. Fischer, J.~Luecker, {Propagators and phase structure of Nf=2 and Nf=2+1
  QCD}, Phys.Lett. B718 (2013) 1036--1043.
\newblock \href {http://arxiv.org/abs/1206.5191} {\path{arXiv:1206.5191}},
  \href {http://dx.doi.org/10.1016/j.physletb.2012.11.054}
  {\path{doi:10.1016/j.physletb.2012.11.054}}.

\bibitem{Fischer:2014ata}
C.~S. Fischer, J.~Luecker, C.~A. Welzbacher, {Phase structure of three and four
  flavor QCD}, Phys.Rev. D90 (2014) 034022.
\newblock \href {http://arxiv.org/abs/1405.4762} {\path{arXiv:1405.4762}},
  \href {http://dx.doi.org/10.1103/PhysRevD.90.034022}
  {\path{doi:10.1103/PhysRevD.90.034022}}.

\bibitem{Borsanyi:2013bia}
S.~Borsanyi, Z.~Fodor, C.~Hoelbling, S.~D. Katz, S.~Krieg, et~al., {Full result
  for the QCD equation of state with 2+1 flavors}, Phys.Lett. B730 (2014)
  99--104.
\newblock \href {http://arxiv.org/abs/1309.5258} {\path{arXiv:1309.5258}},
  \href {http://dx.doi.org/10.1016/j.physletb.2014.01.007}
  {\path{doi:10.1016/j.physletb.2014.01.007}}.

\bibitem{Bazavov:2014noa}
A.~Bazavov, T.~Bhattacharya, C.~DeTar, H.~T. Ding, S.~Gottlieb, et~al., {The
  equation of state in (2+1)-flavor QCD}\href {http://arxiv.org/abs/1407.6387}
  {\path{arXiv:1407.6387}}.

\bibitem{Stoecker:2004qu}
H.~Stoecker, {Collective flow signals the quark gluon plasma}, Nucl. Phys. A750
  (2005) 121--147.
\newblock \href {http://arxiv.org/abs/nucl-th/0406018}
  {\path{arXiv:nucl-th/0406018}}, \href
  {http://dx.doi.org/10.1016/j.nuclphysa.2004.12.074}
  {\path{doi:10.1016/j.nuclphysa.2004.12.074}}.

\bibitem{Steinheimer:2014pfa}
J.~Steinheimer, J.~Auvinen, H.~Petersen, M.~Bleicher, H.~Stöcker, {Examination
  of directed flow as a signal for a phase transition in relativistic nuclear
  collisions}, Phys. Rev. C89~(5) (2014) 054913.
\newblock \href {http://arxiv.org/abs/1402.7236} {\path{arXiv:1402.7236}},
  \href {http://dx.doi.org/10.1103/PhysRevC.89.054913}
  {\path{doi:10.1103/PhysRevC.89.054913}}.

\bibitem{Stephanov:2008qz}
M.~Stephanov, {Non-Gaussian fluctuations near the QCD critical point},
  Phys.Rev.Lett. 102 (2009) 032301.
\newblock \href {http://arxiv.org/abs/0809.3450} {\path{arXiv:0809.3450}},
  \href {http://dx.doi.org/10.1103/PhysRevLett.102.032301}
  {\path{doi:10.1103/PhysRevLett.102.032301}}.

\bibitem{Hohenberg:1977ym}
P.~C. Hohenberg, B.~I. Halperin, {Theory of Dynamic Critical Phenomena}, Rev.
  Mod. Phys. 49 (1977) 435--479.
\newblock \href {http://dx.doi.org/10.1103/RevModPhys.49.435}
  {\path{doi:10.1103/RevModPhys.49.435}}.

\bibitem{Berdnikov:1999ph}
B.~Berdnikov, K.~Rajagopal, {Slowing out-of-equilibrium near the QCD critical
  point}, Phys.Rev. D61 (2000) 105017.
\newblock \href {http://arxiv.org/abs/hep-ph/9912274}
  {\path{arXiv:hep-ph/9912274}}, \href
  {http://dx.doi.org/10.1103/PhysRevD.61.105017}
  {\path{doi:10.1103/PhysRevD.61.105017}}.

\bibitem{Skokov:2012ds}
V.~Skokov, B.~Friman, K.~Redlich, {Volume Fluctuations and Higher Order
  Cumulants of the Net Baryon Number}, Phys. Rev. C88 (2013) 034911.
\newblock \href {http://arxiv.org/abs/1205.4756} {\path{arXiv:1205.4756}},
  \href {http://dx.doi.org/10.1103/PhysRevC.88.034911}
  {\path{doi:10.1103/PhysRevC.88.034911}}.

\bibitem{Csernai:1995zn}
L.~Csernai, I.~Mishustin, {Fast hadronization of supercooled quark - gluon
  plasma}, Phys.Rev.Lett. 74 (1995) 5005--5008.
\newblock \href {http://dx.doi.org/10.1103/PhysRevLett.74.5005}
  {\path{doi:10.1103/PhysRevLett.74.5005}}.

\bibitem{Mishustin:1998eq}
I.~Mishustin, {Nonequilibrium phase transition in rapidly expanding QCD
  matter}, Phys.Rev.Lett. 82 (1999) 4779--4782.
\newblock \href {http://arxiv.org/abs/hep-ph/9811307}
  {\path{arXiv:hep-ph/9811307}}, \href
  {http://dx.doi.org/10.1103/PhysRevLett.82.4779}
  {\path{doi:10.1103/PhysRevLett.82.4779}}.

\bibitem{Scavenius:2000bb}
O.~Scavenius, A.~Dumitru, E.~S. Fraga, J.~T. Lenaghan, A.~D. Jackson, {First
  order chiral phase transition in high-energy collisions: Can nucleation
  prevent spinodal decomposition?}, Phys. Rev. D63 (2001) 116003.
\newblock \href {http://arxiv.org/abs/hep-ph/0009171}
  {\path{arXiv:hep-ph/0009171}}, \href
  {http://dx.doi.org/10.1103/PhysRevD.63.116003}
  {\path{doi:10.1103/PhysRevD.63.116003}}.

\bibitem{Keranen:2002sw}
A.~Keranen, J.~Manninen, L.~Csernai, V.~Magas, {Statistical hadronization of
  supercooled quark gluon plasma}, Phys.Rev. C67 (2003) 034905.
\newblock \href {http://arxiv.org/abs/nucl-th/0205019}
  {\path{arXiv:nucl-th/0205019}}, \href
  {http://dx.doi.org/10.1103/PhysRevC.67.034905}
  {\path{doi:10.1103/PhysRevC.67.034905}}.

\bibitem{Randrup:2009gp}
J.~Randrup, {Phase transition dynamics for baryon-dense matter}, Phys.Rev. C79
  (2009) 054911.
\newblock \href {http://arxiv.org/abs/0903.4736} {\path{arXiv:0903.4736}},
  \href {http://dx.doi.org/10.1103/PhysRevC.79.054911}
  {\path{doi:10.1103/PhysRevC.79.054911}}.

\bibitem{Randrup:2010ax}
J.~Randrup, {Spinodal phase separation in relativistic nuclear collisions},
  Phys.Rev. C82 (2010) 034902.
\newblock \href {http://arxiv.org/abs/1007.1448} {\path{arXiv:1007.1448}},
  \href {http://dx.doi.org/10.1103/PhysRevC.82.034902}
  {\path{doi:10.1103/PhysRevC.82.034902}}.

\bibitem{Steinheimer:2012gc}
J.~Steinheimer, J.~Randrup, {Spinodal amplification of density fluctuations in
  fluid-dynamical simulations of relativistic nuclear collisions},
  Phys.Rev.Lett. 109 (2012) 212301.
\newblock \href {http://arxiv.org/abs/1209.2462} {\path{arXiv:1209.2462}},
  \href {http://dx.doi.org/10.1103/PhysRevLett.109.212301}
  {\path{doi:10.1103/PhysRevLett.109.212301}}.

\bibitem{Herold:2013qda}
C.~Herold, M.~Nahrgang, I.~Mishustin, M.~Bleicher, {Formation of droplets with
  high baryon density at the QCD phase transition in expanding matter},
  Nucl.Phys. A925 (2014) 14--24.
\newblock \href {http://arxiv.org/abs/1304.5372} {\path{arXiv:1304.5372}},
  \href {http://dx.doi.org/10.1016/j.nuclphysa.2014.01.010}
  {\path{doi:10.1016/j.nuclphysa.2014.01.010}}.

\bibitem{Stephanov:1998dy}
M.~A. Stephanov, K.~Rajagopal, E.~V. Shuryak, {Signatures of the tricritical
  point in QCD}, Phys.Rev.Lett. 81 (1998) 4816--4819.
\newblock \href {http://arxiv.org/abs/hep-ph/9806219}
  {\path{arXiv:hep-ph/9806219}}, \href
  {http://dx.doi.org/10.1103/PhysRevLett.81.4816}
  {\path{doi:10.1103/PhysRevLett.81.4816}}.

\bibitem{Stephanov:1999zu}
M.~A. Stephanov, K.~Rajagopal, E.~V. Shuryak, {Event-by-event fluctuations in
  heavy ion collisions and the QCD critical point}, Phys.Rev. D60 (1999)
  114028.
\newblock \href {http://arxiv.org/abs/hep-ph/9903292}
  {\path{arXiv:hep-ph/9903292}}, \href
  {http://dx.doi.org/10.1103/PhysRevD.60.114028}
  {\path{doi:10.1103/PhysRevD.60.114028}}.

\bibitem{Athanasiou:2010kw}
C.~Athanasiou, K.~Rajagopal, M.~Stephanov, {Using Higher Moments of
  Fluctuations and their Ratios in the Search for the QCD Critical Point},
  Phys. Rev. D82 (2010) 074008.
\newblock \href {http://arxiv.org/abs/1006.4636} {\path{arXiv:1006.4636}},
  \href {http://dx.doi.org/10.1103/PhysRevD.82.074008}
  {\path{doi:10.1103/PhysRevD.82.074008}}.

\bibitem{Luo:2015ewa}
X.~Luo, {Energy Dependence of Moments of Net-Proton and Net-Charge Multiplicity
  Distributions at STAR}, PoS CPOD2014 (2015) 019.
\newblock \href {http://arxiv.org/abs/1503.02558} {\path{arXiv:1503.02558}}.

\bibitem{Cleymans:2005xv}
J.~Cleymans, H.~Oeschler, K.~Redlich, S.~Wheaton, {Comparison of chemical
  freeze-out criteria in heavy-ion collisions}, Phys. Rev. C73 (2006) 034905.
\newblock \href {http://arxiv.org/abs/hep-ph/0511094}
  {\path{arXiv:hep-ph/0511094}}, \href
  {http://dx.doi.org/10.1103/PhysRevC.73.034905}
  {\path{doi:10.1103/PhysRevC.73.034905}}.

\bibitem{Jiang:2015hri}
L.~Jiang, P.~Li, H.~Song, {Correlated fluctuations near the QCD critical
  point}\href {http://arxiv.org/abs/1512.06164} {\path{arXiv:1512.06164}}.

\bibitem{Jiang:2015cnt}
L.~Jiang, P.~Li, H.~Song, {Multiplicity fluctuations of net protons on the
  hydrodynamic freeze-out surface}, 2015.
\newblock \href {http://arxiv.org/abs/1512.07373} {\path{arXiv:1512.07373}}.

\bibitem{workinprogress}
M.~Bluhm, M.~Nahrgang, to be published soon.

\bibitem{Bzdak:2012ab}
A.~Bzdak, V.~Koch, {Acceptance corrections to net baryon and net charge
  cumulants}, Phys.Rev. C86 (2012) 044904.
\newblock \href {http://arxiv.org/abs/1206.4286} {\path{arXiv:1206.4286}},
  \href {http://dx.doi.org/10.1103/PhysRevC.86.044904}
  {\path{doi:10.1103/PhysRevC.86.044904}}.

\bibitem{Bzdak:2013pha}
A.~Bzdak, V.~Koch, {Local Efficiency Corrections to Higher Order
  Cumulants}\href {http://arxiv.org/abs/1312.4574} {\path{arXiv:1312.4574}}.

\bibitem{Kitazawa:2011wh}
M.~Kitazawa, M.~Asakawa, {Revealing baryon number fluctuations from proton
  number fluctuations in relativistic heavy ion collisions}, Phys.Rev. C85
  (2012) 021901.
\newblock \href {http://arxiv.org/abs/1107.2755} {\path{arXiv:1107.2755}},
  \href {http://dx.doi.org/10.1103/PhysRevC.85.021901}
  {\path{doi:10.1103/PhysRevC.85.021901}}.

\bibitem{Kitazawa:2012at}
M.~Kitazawa, M.~Asakawa, {Relation between baryon number fluctuations and
  experimentally observed proton number fluctuations in relativistic heavy ion
  collisions}, Phys.Rev. C86 (2012) 024904.
\newblock \href {http://arxiv.org/abs/1205.3292} {\path{arXiv:1205.3292}},
  \href {http://dx.doi.org/10.1103/PhysRevC.86.024904,
  10.1103/PhysRevC.86.069902} {\path{doi:10.1103/PhysRevC.86.024904,
  10.1103/PhysRevC.86.069902}}.

\bibitem{Nahrgang:2014fza}
M.~Nahrgang, M.~Bluhm, P.~Alba, R.~Bellwied, C.~Ratti, {Impact of resonance
  regeneration and decay on the net-proton fluctuations in a hadron resonance
  gas}, Eur. Phys. J. C75~(12) (2015) 573.
\newblock \href {http://arxiv.org/abs/1402.1238} {\path{arXiv:1402.1238}},
  \href {http://dx.doi.org/10.1140/epjc/s10052-015-3775-0}
  {\path{doi:10.1140/epjc/s10052-015-3775-0}}.

\bibitem{Sangaline:2015bma}
E.~Sangaline, {Strongly Intensive Cumulants: Fluctuation Measures for Systems
  With Incompletely Constrained Volumes}\href {http://arxiv.org/abs/1505.00261}
  {\path{arXiv:1505.00261}}.

\bibitem{Gorenstein:2011vq}
M.~I. Gorenstein, M.~Gazdzicki, {Strongly Intensive Quantities}, Phys. Rev. C84
  (2011) 014904.
\newblock \href {http://arxiv.org/abs/1101.4865} {\path{arXiv:1101.4865}},
  \href {http://dx.doi.org/10.1103/PhysRevC.84.014904}
  {\path{doi:10.1103/PhysRevC.84.014904}}.

\bibitem{Schuster:2009jv}
M.~Nahrgang, T.~Schuster, M.~Mitrovski, R.~Stock, M.~Bleicher, {Net-baryon-,
  net-proton-, and net-charge kurtosis in heavy-ion collisions within a
  relativistic transport approach}, Eur.Phys.J. C72 (2012) 2143.
\newblock \href {http://arxiv.org/abs/0903.2911} {\path{arXiv:0903.2911}},
  \href {http://dx.doi.org/10.1140/epjc/s10052-012-2143-6}
  {\path{doi:10.1140/epjc/s10052-012-2143-6}}.

\bibitem{Bzdak:2012an}
A.~Bzdak, V.~Koch, V.~Skokov, {Baryon number conservation and the cumulants of
  the net proton distribution}, Phys.Rev. C87 (2013) 014901.
\newblock \href {http://arxiv.org/abs/1203.4529} {\path{arXiv:1203.4529}},
  \href {http://dx.doi.org/10.1103/PhysRevC.87.014901}
  {\path{doi:10.1103/PhysRevC.87.014901}}.

\bibitem{Hippert:2015rwa}
M.~Hippert, E.~S. Fraga, E.~M. Santos, {Critical versus spurious fluctuations
  in the search for the QCD critical point}\href
  {http://arxiv.org/abs/1507.04764} {\path{arXiv:1507.04764}}.

\bibitem{Mukherjee:2015swa}
S.~Mukherjee, R.~Venugopalan, Y.~Yin, {Real time evolution of non-Gaussian
  cumulants in the QCD critical regime}, Phys. Rev. C92~(3) (2015) 034912.
\newblock \href {http://arxiv.org/abs/1506.00645} {\path{arXiv:1506.00645}},
  \href {http://dx.doi.org/10.1103/PhysRevC.92.034912}
  {\path{doi:10.1103/PhysRevC.92.034912}}.

\bibitem{Nahrgang:2011mg}
M.~Nahrgang, S.~Leupold, C.~Herold, M.~Bleicher, {Nonequilibrium chiral fluid
  dynamics including dissipation and noise}, Phys.Rev. C84 (2011) 024912.
\newblock \href {http://arxiv.org/abs/1105.0622} {\path{arXiv:1105.0622}},
  \href {http://dx.doi.org/10.1103/PhysRevC.84.024912}
  {\path{doi:10.1103/PhysRevC.84.024912}}.

\bibitem{Nahrgang:2011mv}
M.~Nahrgang, S.~Leupold, M.~Bleicher, {Equilibration and relaxation times at
  the chiral phase transition including reheating}, Phys.Lett. B711 (2012)
  109--116.
\newblock \href {http://arxiv.org/abs/1105.1396} {\path{arXiv:1105.1396}},
  \href {http://dx.doi.org/10.1016/j.physletb.2012.03.059}
  {\path{doi:10.1016/j.physletb.2012.03.059}}.

\bibitem{Nahrgang:2011vn}
M.~Nahrgang, C.~Herold, S.~Leupold, I.~Mishustin, M.~Bleicher, {The impact of
  dissipation and noise on fluctuations in chiral fluid dynamics}, J.Phys. G40
  (2013) 055108.
\newblock \href {http://arxiv.org/abs/1105.1962} {\path{arXiv:1105.1962}},
  \href {http://dx.doi.org/10.1088/0954-3899/40/5/055108}
  {\path{doi:10.1088/0954-3899/40/5/055108}}.

\bibitem{Herold:2013bi}
C.~Herold, M.~Nahrgang, I.~Mishustin, M.~Bleicher, {Chiral fluid dynamics with
  explicit propagation of the Polyakov loop}, Phys.Rev. C87 (2013) 014907.
\newblock \href {http://arxiv.org/abs/1301.1214} {\path{arXiv:1301.1214}},
  \href {http://dx.doi.org/10.1103/PhysRevC.87.014907}
  {\path{doi:10.1103/PhysRevC.87.014907}}.

\bibitem{Herold:2014zoa}
C.~Herold, M.~Nahrgang, Y.~Yan, C.~Kobdaj, {Net-baryon number variance and
  kurtosis within nonequilibrium chiral fluid dynamics}, J.Phys. G41~(11)
  (2014) 115106.
\newblock \href {http://arxiv.org/abs/1407.8277} {\path{arXiv:1407.8277}},
  \href {http://dx.doi.org/10.1088/0954-3899/41/11/115106}
  {\path{doi:10.1088/0954-3899/41/11/115106}}.

\bibitem{Herold:2016uvv}
C.~Herold, M.~Nahrgang, Y.~Yan, C.~Kobdaj, {Dynamical net-proton fluctuations
  near a QCD critical point }\href {http://arxiv.org/abs/1601.04839}
  {\path{arXiv:1601.04839}}.

\bibitem{Son:2004iv}
D.~Son, M.~Stephanov, {Dynamic universality class of the QCD critical point},
  Phys.Rev. D70 (2004) 056001.
\newblock \href {http://arxiv.org/abs/hep-ph/0401052}
  {\path{arXiv:hep-ph/0401052}}, \href
  {http://dx.doi.org/10.1103/PhysRevD.70.056001}
  {\path{doi:10.1103/PhysRevD.70.056001}}.

\bibitem{Kapusta:2011gt}
J.~I. Kapusta, B.~Muller, M.~Stephanov, {Relativistic Theory of Hydrodynamic
  Fluctuations with Applications to Heavy Ion Collisions}, Phys. Rev. C85
  (2012) 054906.
\newblock \href {http://arxiv.org/abs/1112.6405} {\path{arXiv:1112.6405}},
  \href {http://dx.doi.org/10.1103/PhysRevC.85.054906}
  {\path{doi:10.1103/PhysRevC.85.054906}}.

\bibitem{Murase:2013tma}
K.~Murase, T.~Hirano, {Relativistic fluctuating hydrodynamics with memory
  functions and colored noises}\href {http://arxiv.org/abs/1304.3243}
  {\path{arXiv:1304.3243}}.

\bibitem{Kapusta:2012zb}
J.~I. Kapusta, J.~M. Torres-Rincon, {Thermal Conductivity and Chiral Critical
  Point in Heavy Ion Collisions}, Phys. Rev. C86 (2012) 054911.
\newblock \href {http://arxiv.org/abs/1209.0675} {\path{arXiv:1209.0675}},
  \href {http://dx.doi.org/10.1103/PhysRevC.86.054911}
  {\path{doi:10.1103/PhysRevC.86.054911}}.

\bibitem{Young:2014pka}
C.~Young, J.~I. Kapusta, C.~Gale, S.~Jeon, B.~Schenke, {Thermally Fluctuating
  Second-Order Viscous Hydrodynamics and Heavy-Ion Collisions}, Phys. Rev.
  C91~(4) (2015) 044901.
\newblock \href {http://arxiv.org/abs/1407.1077} {\path{arXiv:1407.1077}},
  \href {http://dx.doi.org/10.1103/PhysRevC.91.044901}
  {\path{doi:10.1103/PhysRevC.91.044901}}.

\bibitem{Chafin:2012eq}
C.~Chafin, T.~Schäfer, {Hydrodynamic fluctuations and the minimum shear
  viscosity of the dilute Fermi gas at unitarity}, Phys. Rev. A87~(2) (2013)
  023629.
\newblock \href {http://arxiv.org/abs/1209.1006} {\path{arXiv:1209.1006}},
  \href {http://dx.doi.org/10.1103/PhysRevA.87.023629}
  {\path{doi:10.1103/PhysRevA.87.023629}}.

\bibitem{Kovtun:2011np}
P.~Kovtun, G.~D. Moore, P.~Romatschke, {The stickiness of sound: An absolute
  lower limit on viscosity and the breakdown of second order relativistic
  hydrodynamics}, Phys. Rev. D84 (2011) 025006.
\newblock \href {http://arxiv.org/abs/1104.1586} {\path{arXiv:1104.1586}},
  \href {http://dx.doi.org/10.1103/PhysRevD.84.025006}
  {\path{doi:10.1103/PhysRevD.84.025006}}.

\end{thebibliography}
\bibliographystyle{elsarticle-num}







\end{document}